\documentclass[]{iopart}
 
\usepackage{graphicx}
\usepackage{amssymb}
\usepackage{latexsym}
\usepackage{epsfig}

\newcommand{\dii}{{\mathrm d}}
\newcommand{\ima}{{\mathrm i}}

\newcommand{\vex}{\vec{x}}

\newcommand{\veq}{\vec{q}}

\newcommand{\LAM}{{\bf \Lambda}}

\newcommand{\ii}{{\rm i}}

\newcommand{\Laq}{\lambda_Q}
\newcommand{\ranGC}{\rangle_{\mathrm{GC}}}

\newcommand{\ch}{{\mathrm{ch}}}

\begin{document}

\title[Particle Number Fluctuations and 
Exact Charge Conservation Laws]
{Particle Number Fluctuations in Statistical Model
with Exact Charge Conservation Laws}

\date{}
\author{A. Ker\"anen\dag\footnote{Electronic address: 
antti.keranen@oulu.fi}, F. Becattini\dag,
 V.V.~Begun$\|$, M.I.~Gorenstein$\|$\ and O.S.~Zozulya$\|$\S}
\address{\dag\ INFN Sezione di Firenze, Italy}
\address{$\|$\ Bogolyubov Institute for Theoretical Physics, Kiev, Ukraine}
\address{\S\ Utrecht University, Utrecht, The Netherlands}


\begin{abstract} 
Even though the first momenta i.e. the ensemble average quantities in
canonical ensemble (CE) give the grand canonical (GC)
results in large multiplicity limit, the fluctuations involving second
momenta do not respect this asymptotic behaviour.
Instead, the asymptotics are strikingly different, giving a new handle
in study of statistical particle number fluctuations in
relativistic nuclear reactions.
Here we study the analytical large volume asymptotics to
general case of multispecies hadron gas carrying fixed baryon number,
strangeness and electric charge.
By means of Monte Carlo simulations  we have also studied the general
multiplicity probability distributions taking into account the
decay chains of resonance states. 
\end{abstract}

\vspace{-0.7cm}
\section{Introduction}
\vspace{-0.2cm}

The surprising success of the statistical hadronization model
in describing the multihadron production in high-energy nucleus-nucleus
reactions (see e.g. \cite{thermal}
 and references therein) as well as in 
elementary particle collisions \cite{elementary}
gives rise to a question
if the fluctuations, especially the second moments of 
multiplicity distributions,
 follow the statistical resonance gas results
 as well as the mean multiplicities
(the first momenta of multiplicities).
Even though the CE first momenta reach the GC values quite fast with increasing
multiplicities \cite{kerabeca}, we have found that
 the variance of these values {\em always} stay different
in the CE \cite{gorega,begun}
due to the lack of charge bath,
 i.e. due to the lack of fluctuations of the
relevant net charges. 

In section \ref{theory} we pose and solve the theoretical problems involved,
and give results for the general hadron gas.
 The section \ref{numerical} is devoted to the numerical
studies employing the methods developed in previous section, and
to the comparison of those with the ones given by the Monte Carlo (MC) 
procedure explained
in Ref. \cite{MC1}.

\section{Theoretical Studies}\label{theory}
\vspace{-0.2cm}

The central quantity in our discussion is the {\em scaled variance}
of the multiplicity $N$:
\begin{equation} \label{eq:omega}
\omega = \frac{\langle N^2 \rangle -\langle N \rangle^2}{\langle N \rangle}.
\end{equation}
This is a finite measurable parameter
 even at the infinite volume limit, and we will
show it losing its volume dependence in large systems.
It is worth stressing that for the Poissonian distribution corresponding
to Boltzmann GC ensemble, $\omega = 1$. Although we keep the 
theoretical discussion on the level of Boltzmann approximation for brevity,
we show also 
the full quantum statistics results in the section \ref{numerical}. 

In the CE, the partition function does not factorize to
one-species parts, but the different species are bound by the
group factors giving delta functions
over corresponding exact conservation laws.
Recall the form of the partition function for canonical hadron gas
carrying baryon number $B$, strangeness $S$ and electric charge $Q$
\cite{projection}: \vspace{-0.3cm}
\begin{equation}
Z_{\vec{Q}}(T,V) = \left[ \prod_{i=1}^{3}
\frac{1}{2\pi}  \int_0^{2\pi}\dii\phi_i e^{-\ii Q_i\phi_i}\right]
         Z_{GC}(T,V,\{\lambda_{Q_i}\}),
\label{eq:projection}
\end{equation}
where $\vec{Q_i} = (Q_1,Q_2,Q_3) = (B,S,Q)$.
$\phi_i \in [0,2\pi)$ is a $U(1)$ group parameter and a Wick-rotated
fugacity factor $\lambda_{Q_i} \rightarrow e^{i\phi_i}$
 is introduced for every charge
in the grand canonical partition function $Z_{GC}.$

By the substitution $x_i = e^{\ii \phi_i}$ and still integrating
over the three unit circles we may write
\begin{small}
\begin{eqnarray}\label{eq:partx}
Z_{\vec{Q}-\veq_i}(T,V)
&=& \frac{1}{(2\pi\ima)^3} \oint \dii x_B \dii x_S \dii x_Q
\ x_B^{B_i-1}x_S^{S_i-1}x_Q^{Q_i-1} \\
&& \times \exp\!\left\{-B\ln x_B-S\ln x_S-Q\ln x_Q
+\sum_j z_j^1 x_B^{B_i}x_S^{S_i}x_Q^{Q_i}\right\}. \nonumber \\
&\equiv& \frac{1}{(2\pi\ima)^3} \oint \dii x_B \dii x_S \dii x_Q \,
g(\vec{x})\times \exp\left\{f(\vex)\right\}.\nonumber
\end{eqnarray}
\end{small}

The first and second momenta of a given multiplicity of a hadron subset
$h$  can be calculated
by inserting fictitious fugacity to the generating function
$Z_{GC}(T,V,\{\lambda_{Q_i}\rightarrow e^{i\phi_i}\})$ in equation
(\ref{eq:projection}) and performing the derivatives
\begin{small}
\begin{eqnarray} \label{eq:momenta}
\langle N_h \rangle &= \left.
\frac{1}{Z_{\vec{Q}}}\sum_{j\in h}\frac{\partial Z_{\vec{Q}}}
{\partial \lambda_j}\right|_{\lambda_j=1}
&= \sum_{j\in h}z_j^1 \frac{Z_{\vec{Q}-\veq_j}}{Z_{\vec{Q}}}\\
\langle N_h^2 \rangle &=
 \frac{1}{Z_{\vec{Q}}}
\sum_{j\in h}\sum_{i\in h}
\left[\frac{\partial}{\partial\lambda_j}\left(\lambda_i
 \frac{\partial Z_{\vec{Q}}}
{\partial \lambda_i}\right)\right]_{\lambda_{j,i}=1}
&= \sum_{j\in h} z_j^1 \frac{Z_{\vec{Q}-\veq_j}}{Z_{\vec{Q}}}
+  \sum_{j\in h}z_j^1\sum_{i\in h}z_i^1
\frac{Z_{\vec{Q}-\veq_j-\veq_i}}{Z_{\vec{Q}}}. \nonumber
\end{eqnarray}
\end{small}
Using these, the scaled variance can be written as the sum of
Poissonian part and the canonical correction:
\begin{equation} \label{eq:MBomega}
\omega_h = 1 + \frac{\sum_{i\in h} \langle N_i \rangle
\sum_{j\in h} z^1_j\left( \frac{Z_{\vec{Q}-\vec{q}_i-\vec{q}_j}}
{Z_{\vec{Q}-\vec{q}_i}}
- \frac{Z_{\vec{Q}-\vec{q}_j}}{Z_{\vec{Q}}} \right)}
{\sum_{i\in h} \langle N_i \rangle}.
\end{equation}

For the large systems, the canonical correction factors can be solved by
means of the asymptotic expansion, where the saddlepoint of the $f(\vex)$
(\ref{eq:partx})
coincides with the values of GC fugacities \cite{asy}. 
In the general case of multi-hadron gas carrying total charges
$\{Q_k\} = (B,S,Q)$ the asymptotic expansion of the partition function
involves diagonalization of the Hessian matrix
\begin{small}
\begin{equation}
H_{Q_kQ_l}(\vec{x}_0) = \left. \frac{\partial^2 f(\vex)}
{\partial x_{Q_k}\partial x_{Q_l}}\right|_{\vex_0}
= \lambda_{Q_k}^{-1}\lambda_{Q_l}^{-1}
\left\{Q_k\delta_{kl} + \sum_j (Q_k)_j\left[(Q_l)_j-\delta_{kl}\right]
\langle N_j \rangle_{\mathrm{GC}}\right\}.
\end{equation}
\end{small}
Because this is real and symmetric, the diagonalization is done
by the orthogonal matrix $\LAM:$
$
H' = {\mathrm{diag}}(h_1,h_2,h_3) = \LAM^T H \LAM,
$
where $h_k$ are the Hessian eigenvalues, and $\LAM$ is constructed
of corresponding eigenvectors.
Applying these and the notation $\langle N_i \ranGC$ for GC average 
multiplicities, the large volume limit of the Eq. (\ref{eq:MBomega})
is found to be
\begin{small}
\begin{equation}\label{eq:canom}
\omega_h = 1 - \frac{\sum_{i\in h} \langle N_i \ranGC
\sum_{j\in h} \langle N_j \ranGC}
{\sum_{i\in h} \langle N_i \ranGC} \sum_{Q_k}\sum_{Q_m}(Q_k)_i(Q_m)_j
\lambda_{Q_k}\lambda_{Q_m}\sum_l\frac{\bar{\Lambda}^l_{Q_k}
\bar{\Lambda}^l_{Q_m}}{h_l},
\end{equation}
\end{small}
where the summations of $Q_{k,l}$ are over relevant charges
$B$, $S$ and $Q$, and the $ \bar{\Lambda}^l_{Q_k}$ are the
elements of the transformation matrix $\LAM^T$.

As an easily evaluable approximation, we consider the scaled variance for
the number of positively ($h^+$) and negatively ($h^-$)
 charged particles as well as
the sum of all charged particles neglecting the multi-charged particles
and the mixing between different charges (non-diagonals in the Hessian).
The relevant Hessian term is
$H_{QQ} = \Laq^{-2}\left[Q+\sum_j z_1^j \lambda_j Q_j(Q_j-1)\right]$,
which leads to $H_{QQ} = \Laq^{-2}\left[Q+ 2\langle h^- \ranGC\right]
= \Laq^{-2}\left[\langle h^+ \ranGC +\langle h^- \ranGC\right]$.
This yields the terms
$Q_iQ_j/(\langle h^+ \ranGC +\langle h^- \ranGC)$ inside the $Q_{k,m}$
summations in the Eq. (\ref{eq:canom}).
Applying this to equation (\ref{eq:canom}) yields
\begin{eqnarray} \label{eq:pmomega}
\lim_{V\rightarrow\infty}\omega_\pm \simeq 1-\frac{\langle h^\pm \ranGC}
{\langle h^+ \ranGC + \langle h^- \ranGC} ;\ \ 
\lim_{V\rightarrow\infty}\omega_\ch \simeq 1-\left(
\frac{\langle h^+ \ranGC - \langle h^- \ranGC}
{\langle h^+ \ranGC + \langle h^- \ranGC}\right)^2. 
\end{eqnarray}
This result serves as a ``rule of thumb'' for any charge under consideration
-- it applies trivially to baryons and strange particles too -- but
in quantitative comparisons one must relax the approximations used above.
In case of one-charge system, such as pion gas, Eq. (\ref{eq:pmomega})
is an exact result for thermodynamical limit.
One sees immediately, that for a neutral system 
$\lim_{V\rightarrow\infty}\omega_\pm = 0.5$ instead of GC limit
$\omega_\pm = 1$. When charge density grows indefinitely, the limiting values
are  $\lim_{V\rightarrow\infty}\omega_+ = 0$, 
$\lim_{V\rightarrow\infty}\omega_- = 1$ and 
$\lim_{V\rightarrow\infty}\omega_\ch = 0$. 

\section{Numerical Results}\label{numerical}
\vspace{-0.2cm}

In some special cases, Eq. (\ref{eq:MBomega}) can be easily evaluated
for any volume \cite{gorega,begun}, not only in the asymptotic limit.
Thus, in Fig. 1 we quote the neutral pion gas charged particle
 fluctuations as functions of GC multiplicities \cite{gorega}. 
Also in Fig. 1, we show the asymptotic large volume
results as functions of baryochemical potential
in the nucleon-pion gas in $BQ$-canonical
calculation \cite{begun} together with the 
approximation where the $B$ is handled in 
GC manner, see Eq. (\ref{eq:pmomega}).
\begin{figure}[t]
\begin{center}
$\begin{array}{c@{\hspace{1in}}c}
    \epsfxsize=2.7in
    \epsffile{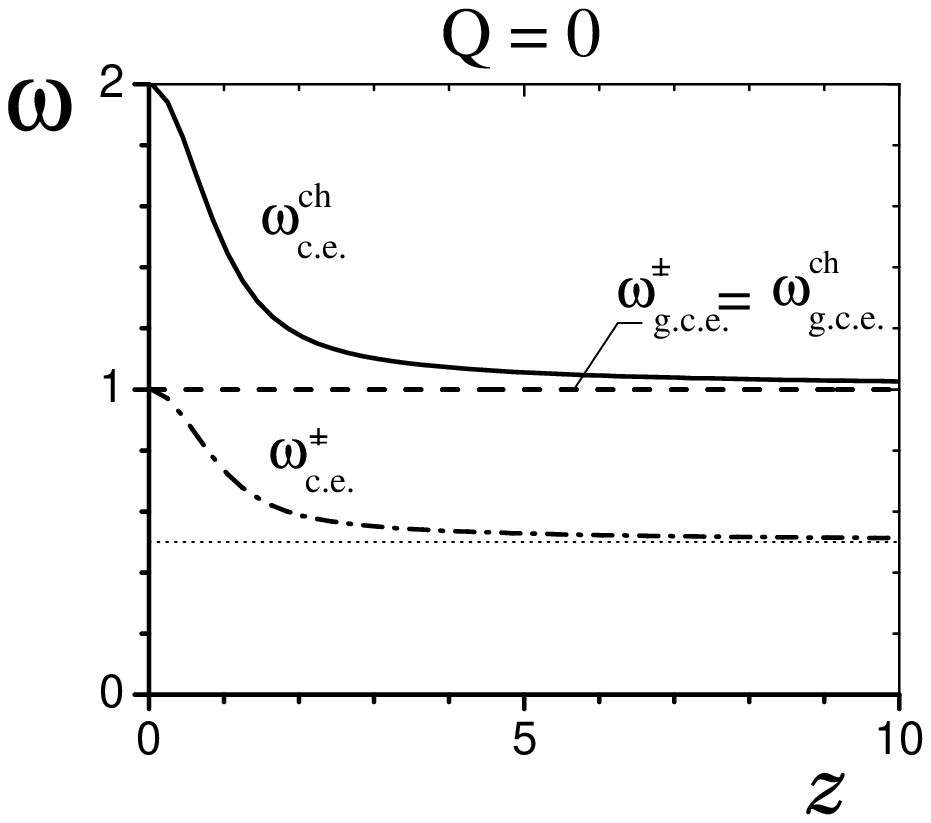} & \hspace*{-3.7cm}
        \epsfxsize=2.8in
        \epsffile{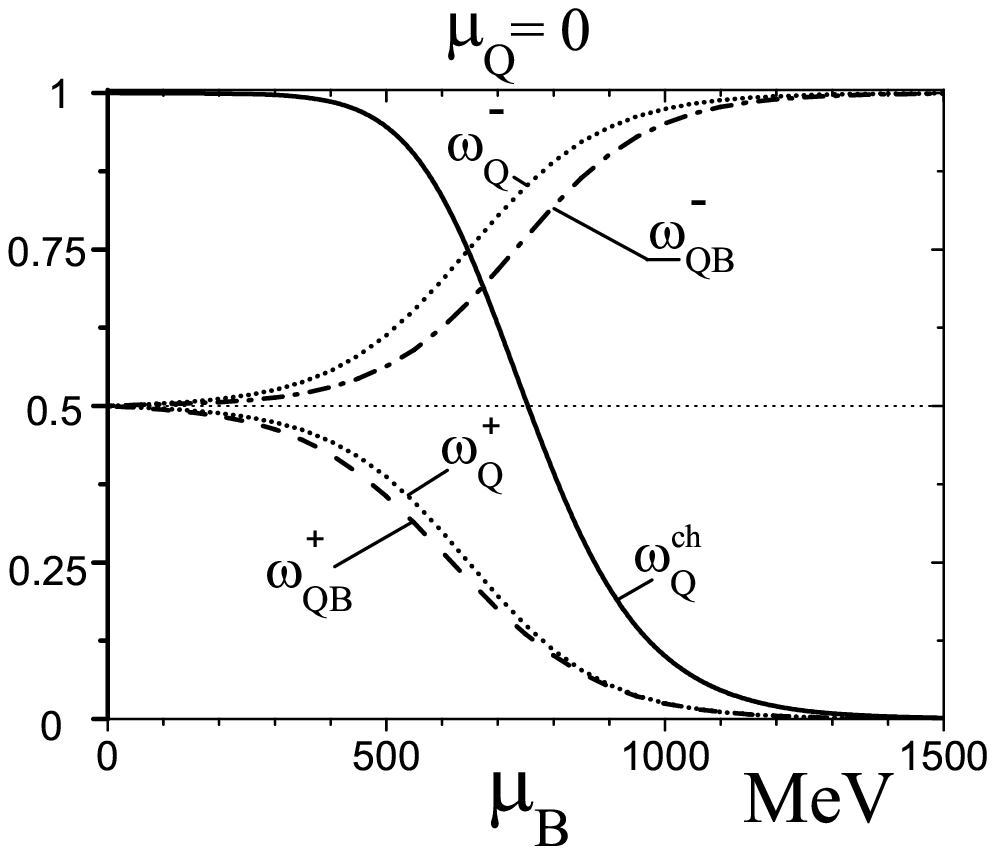} \\
\end{array}$
\vspace{-0.7cm}
\end{center}
\caption{LEFT: Scaled variances in the canonical (c.e.) 
neutral pion gas as functions
of GC one-particle partition functions (i.e. GC multiplicities).
GC Poissonian value is denoted by g.c.e.
\\
RIGHT: Scaled variances in the thermodynamical limit in
 nucleon-pion gas ($T=120$ MeV) as functions of 
baryochemical potential $\mu_B$.
 $\omega_{QB}$
stands for $QB$ canonical result, whereas for $\omega_Q$ the GC
approximation is used for baryon number conservation.
In the former case, $\mu_B$ is taken only as a saddle point in the projection
integral (\ref{eq:partx}), not as a Lagrangian coefficient.}
\end{figure}

In Fig. 2 we turn to the full $BSQ$-canonical $\omega$ 
calculations as functions of baryon density for temperature
$T=160$ MeV. The isospin condition is
chosen to correspond to the PbPb and AuAu reactions: $Q/B = 0.4$.
 We take into
account all the resonances up to the mass $\sim 1.9$ GeV.
The left panel shows the asymtotic results for scaled variances of
primary negative, positive and charged hadrons together with
Monte Carlo results for $V=200$ fm$^3$, which is checked to be large enough
to give asymptotic results. Also shown are the values for $\omega_-$ and
$\omega_\ch$ after the strong decay chain.
In the right panel we present the scaled variances for primary
 baryons, antibaryons,
strange hadrons and anti-strange hadrons.
\begin{figure}[t]
\begin{center}
$\begin{array}{c@{\hspace{1in}}c}
    \epsfxsize=3.3in
    \epsffile{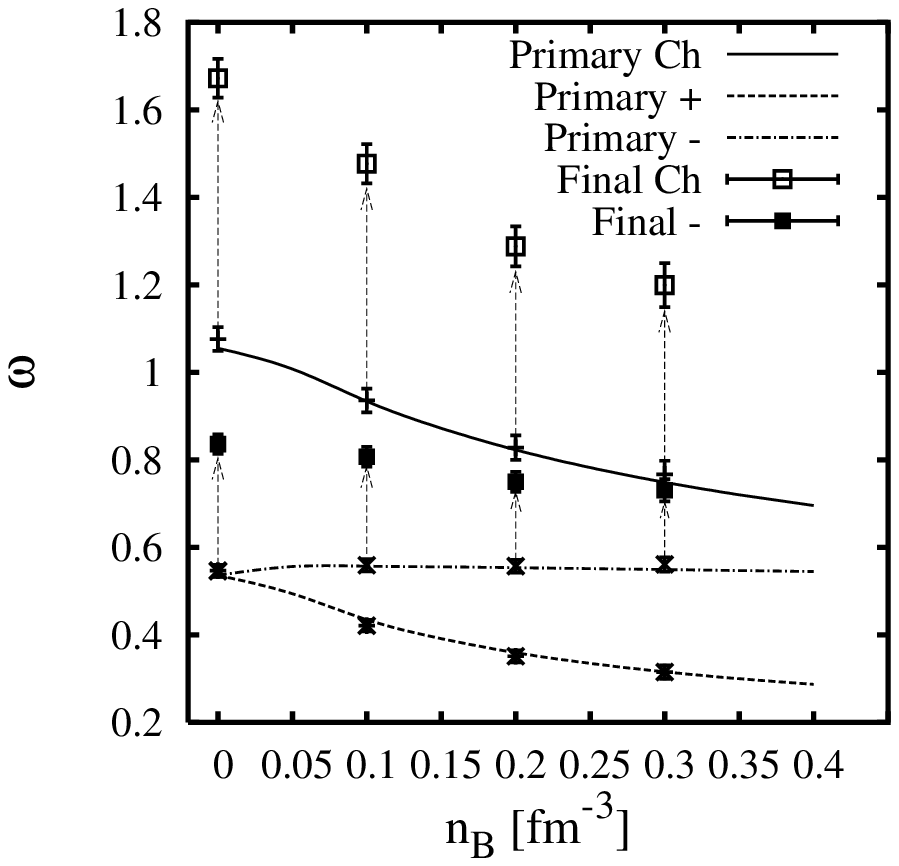} & \hspace*{-4.5cm}
        \epsfxsize=3.3in
        \epsffile{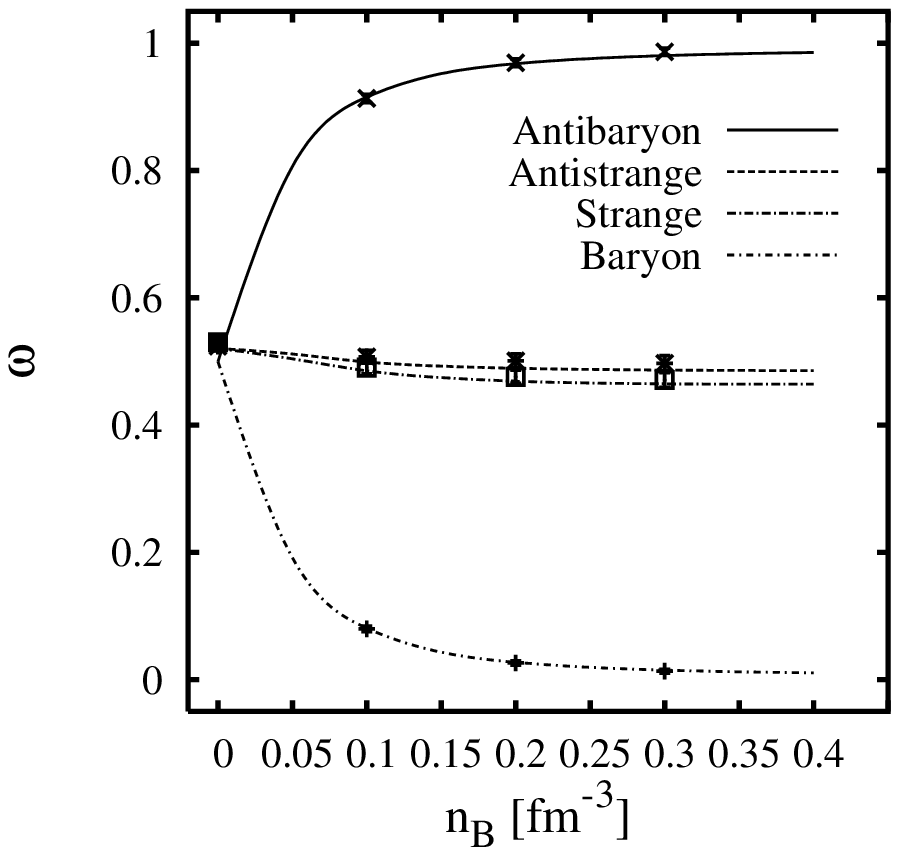} \\
\end{array}$
\vspace{-0.7cm}
\end{center}
\caption{Full quantum hadron gas results in $T=160$ MeV, $Q/B=0.4$.
LEFT: Scaled variances for charged, positive and negative hadrons
as functions of baryon density. The lines depict the asymtotic results,
and dots are for MC results together with statistical errorbars.
The arrows show the change of $\omega_-$ and $\omega_\ch$, when the strong
decay chains are taken into account in addition to primary production.\\
RIGHT: Scaled variances for primary baryons, antibaryons, strange hadrons and 
for antistrange hadrons.}
\end{figure}

\ack
\vspace{-0.2cm}
The authors would like to acknowledge the inspiring and useful discussions
with J.~Cleymans, K.~Redlich and L.~Turko. A.K. and F.B. would like to
 thank the  organizers for the greatly enjoyable conference environment.


\end{document}